\documentclass{mpe_report}

\usepackage{psfig,graphicx,epsfig}
\usepackage{color}
\usepackage{amsmath,amssymb,epic,eepic,array}

\begin{document}

\pagenumbering{arabic}
\setcounter{page}{52}

\renewcommand{\FirstPageOfPaper }{ 52}\renewcommand{\LastPageOfPaper }{ 55}

\title{Pulsar Surveys Present and Future: The Arecibo PALFA Survey and Projected SKA Survey}
\author{J.~S.~Deneva\inst{1}, J.~M.~Cordes, D.~R.~Lorimer, P.~C.~C.~Freire, F.~Camilo, I.~H.~Stairs,
D.~J.~Nice, D.~J.~Champion, J.~van Leeuwen, R.~Ramachandran, A.~J.~Faulkner, A.~G.~Lyne,
S.~M.~Ransom, Z.~ Arzoumanian, R.~ N.~ Manchester, M.~ A.~ McLaughlin,
J.~W.~T.~Hessels, W.~Vlemmings, A.~A.~Deshpande, N.~D.~R.~Bhat, S.~Chatterjee,
J.~L.~Han, B.~M.~Gaensler, L.~Kasian, B.~Reid, T.~J.~W.~Lazio,
V.~M.~Kaspi, F.~Crawford, A.~N.~Lommen, D.~C.~Backer, M.~Kramer, B.~W.~Stappers,
G.~B.~Hobbs, A.~Possenti, N.~D'Amico and C.-A.~Faucher-Gigu\`ere}  
\institute{Cornell University, 516 Space Sciences, Ithaca, NY 14853, USA, \tt{deneva@astro.cornell.edu}
\and \tt{http://www.naic.edu/alfa/pulsar}}
\authorrunning{Deneva, J. S. et al.}
\titlerunning{The Arecibo PALFA Survey and Future SKA Survey}
\maketitle

\begin{abstract}

The Arecibo Pulsar-ALFA (PALFA) survey of the Galactic plane began in 2004 when the new ALFA (Arecibo L-band Feed Array) receiver was commissioned. It is slated to continue for the next 3-5 years and is expected to discover hundreds of new pulsars. We present the goals, progress, and recent discoveries of the PALFA survey. So far preliminary data processing has found 24 new pulsars, one of which is a young 144ms pulsar in a highly relativistic binary with an orbital period of 3.98 hours. Another object exhibits sporadic bursts characteristic of a newly defined class of radio-loud neutron stars: RRATs (Rotating RAdio Transients). The PALFA survey is going to accumulate a total of ~1
Petabyte (PB) of raw data which will be made available to the community via a sophisticated tape archive and database hosted at the Cornell Theory Center supercomputing facility. Web tools and services are being developed which will allow users of the archive to perform data analysis remotely.  

We also discuss parameters, expected discoveries and the scientific impact of a projected pulsar survey with the Square Kilometer Array (SKA). The SKA is to become operational in 2014 and will have the capability of detecting thousands of pulsars not detectable with current instruments, allowing us to perform a comprehensive census of the Galactic pulsar population. 
\end{abstract}

\section{The Arecibo Pulsar-ALFA Survey}
\subsection{Survey parameters}

\begin{figure}
\centerline{\psfig{file=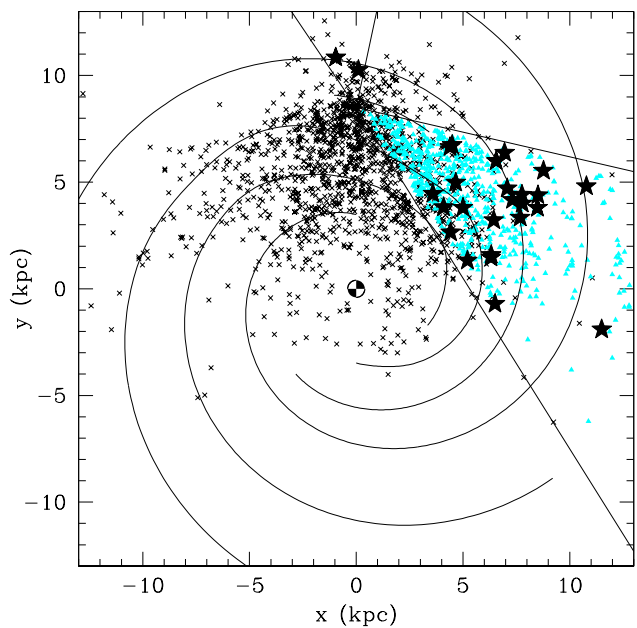,width=8.8cm,clip=}}
\caption{Pulsar locations in the Galactic plane. Crosses denote known pulsars, triangles denote simulated PALFA discoveries, stars denote actual PALFA discoveries. Slanting lines delimit regions of the Galactic plane accessible to the Arecibo telescope. Distances were estimated based on the NE2001 model of the electron density in the Galaxy (Cordes \& Lazio 2002).}
\label{galplane}
\end{figure}      

The PALFA survey of the Galactic plane was enabled by the installation of the new 7-beam ALFA receiver at the Arecibo telescope. It began in 2004 and is slated to continue for 3-5 years, discovering $\sim 1000$ pulsars (Fig.\ref{galplane}). The survey covers Galactic latitudes $|b| \leq 5^\circ$ in the intervals of longitude accessible to the Arecibo telescope, $32^\circ \leq l \leq 77^\circ$ and $168^\circ \leq l \leq 214^\circ$. The ALFA receiver features 6 beams in a hexagonal configuration around a central beam. The half-power beam diameter at 1.42 GHz is $\sim 3$ arcminutes, and the on-axis gain is 10.4 K Jy$^{-1}$ for the central beam and $\sim 8.2$ K Jy$^{-1}$ for the other beams. The system temperature looking out of the Galactic plane is $\sim 24$ K (Heiles 2004, Cordes et al. 2005). In order to cover the sky with gain equal to or exceeding the half-maximum gain, three 7-beam pointings must be closely tiled. Rather than proceed sequentially with tiling the sky, we do sparse sampling, completing one pointing per triple with the intention to fill in the gaps after all pointings from this first sparse grid are completed. The sparse sampling scheme makes use of the large sidelobes of the 6 outer beams which have $\sim 16\%$ of the on-axis gain, thus surveying more sky area per unit time (Cordes et al. 2005 and references therein).    

Currently the PALFA survey uses the four Wideband Arecibo Pulsar Processor (WAPP) backends (Dowd et al. 2000) with a bandwidth of 100 MHz centered on 1.42 GHz, 256 channels, and a sampling time of 64 $\mu$s. In early 2007, we will begin using a new spectrometer capable of processing the full 300 MHz receiver bandwidth with 1024 channels. The data is decimated and processed in quasi-real time during the observations using publicly available ``quicklook'' software (Lorimer 2001). In this pipeline some resolution is traded for speed, and the data is decimated by a factor of 8 in frequency and 16 in time. This means that initial data processing is sensitive mainly to non-recycled pulsars with periods greater than $\sim 30$ ms. Decimated datasets have a time resolution of 1024 $\mu$s and are dedispersed with 96 trial DM values. The resulting time series are then searched for periodic signals using FFTs and harmonic summing, and for single pulses using a matched filtering algorithm (Cordes \& McLaughlin 2003). Processing full-resolution data will be performed on dedicated clusters at the home institutions of PALFA consortium members and will include acceleration searches, which will greatly increase sensitivity to millisecond pulsars and pulsars in short-period binaries. 

\subsection{Results}

\begin{figure}[!h]
\centerline{\psfig{file=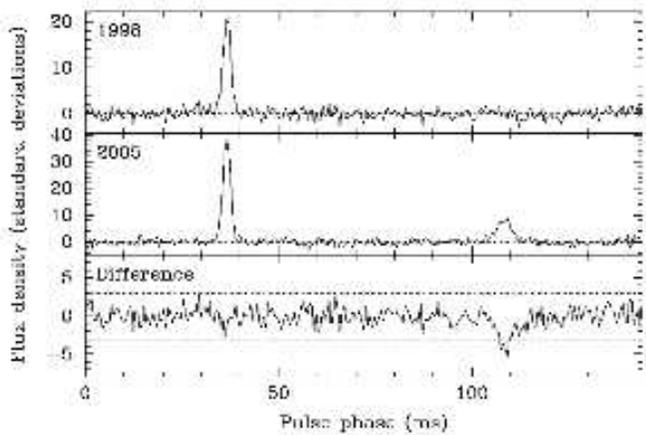,width=8.9cm,clip=}}
\caption{J1906+0746 pulse profiles at 1.374 GHz from a 1998 (top) and 2005 (middle) Parkes observation. The bottom panel shows the difference of the two profiles after scaling them for the area of the main pulse. The dashed horizontal lines delimit $\pm 3\sigma$ calculated from the pulse profile baseline noise (Lorimer et al.2005).}
\label{binary}
\end{figure}

As full-resolution processing pipelines are still being tested, the 28 pulsars discovered so far were found in decimated data only. The highlight among these new objects is J1906+0746, a young, relativistic binary pulsar with a period of 144 ms and DM of 217.78 pc cm$^{-3}$ (Table.\ref{table1}). The discovery observation in September 2004 detected it with a S/N $\sim 11$. The short integration time of the PALFA pointing (134 seconds) did not allow for its binary nature to be immediately seen, but subsequent processing of an archive 35-minute Parkes observation of the same region of the sky revealed that the 144 ms signal was accelerated and follow-up observations revealed its orbital period to be 3.98 hours (Lorimer et al. 2005). Without accounting for acceleration, the pulsar was below the detection threshold of $8-9\sigma$ for the Parkes Multibeam Survey and was rejected as a candidate because of known radio interference close to its period. J1906+0746 has a characteristic age of 112 kyr, and the total system mass is 2.61~$\pm$~0.02~M$_{\odot}$, as indicated by the measured rate of advance of periastron. If we assume the pulsar mass to be close to 1.4~M$_{\odot}$, this leaves leeway for the companion to be either a neutron star or a massive white dwarf. Deep searches for pulsed signals from the companion have not found any plausible candidates. From these results, we can place an upper limit of 0.1 mJy kpc$^2$ on the luminosity of the hypothetical pulsar companion. Another possibility is that the companion may be a pulsar which is not beamed towards us. 
J1906+0746 shows a pulse profile variation between 1998 and 2005 (Fig.\ref{binary}) which can be explained by geodetic precession. With time, the precessing beam of the pulsar changes its orientation with respect to our line of sight and that causes the observed pulse shape to evolve. In this case, an interpulse is not visible in the 1998 observation, but appears in the 2005 data. 

\begin{table}[]
\caption{PSR J1906+0746 parameters (Lorimer et al. 2005).}
\begin{tabular}{lr}
  \noalign{\smallskip}
  \hline
  \noalign{\smallskip}
  Parameter  &  Value \\
  \noalign{\smallskip}
  \hline
  \noalign{\smallskip}            
  Right ascension (J2000) & 19$^h$06$^m$48.$^s$673(6) \\
  Declination (J2000) & 07$^\circ$46'28.6(3)'' \\
  Spin period, $P$ (ms) & 144.071929982(3) \\
  Spin period derivative, $\dot{P}$ & $2.0280(2)\times 10^{-14}$ \\
  Epoch (MJD) & 53590 \\
  Orbital period, $P_b$ (days) & 0.165993045(8) \\
  Projected semi-major axis, $x$ (lt s) & 1.420198(2) \\
  Orbital eccentricity, $e$ & 0.085303(2) \\
  Epoch of periastron, $T_0$ (MJD) & 53553.9126685(6) \\
  Periastron advance rate, $\dot{\omega}$ ($^\circ$ yr$^{-1}$) & 7.57(3) \\
  Dispersion measure, DM (pc cm$^{-3}$) & 217.780(2) \\
  Rotation measure, RM (rad m$^{-2}$) & $+$150(10) \\
  Flux density at 1.4~GHz, $S_{1.4}$ (mJy) & 0.55(15) \\
  Main pulse widths at 50\% and 10\% (ms) & 0.6 and 1.7 \\
  Characteristic age, $\tau_c = \frac{P}{2\dot{P}}$ (kyr) & 112 \\
  Inferred distance, $d$ (kpc) & $\sim 5.4$ \\
  Spectral index, $\alpha$ & -1.3(2) \\
  Mass function, $f$ (M$_\odot$) & 0.1116222(6) \\
  Total system mass, $M_{tot}$ (M$_\odot$) & 2.61(2) \\
  Grav. wave coalescence time, $\tau_g$ (Myr) & $\sim 300$ \\
  \noalign{\smallskip}
  \hline
\label{table1}
\end{tabular} 
\end{table} 

\begin{figure}[!h]
\centerline{\psfig{file=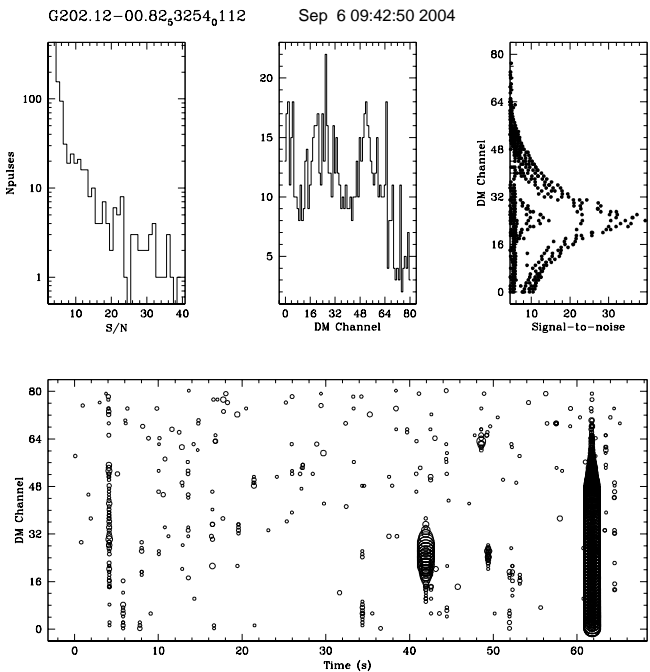,width=9.5cm,clip=}}
\caption{Discovery observation of RRAT J0628+09. Panels on top show total number of events vs. S/N (left), number of events vs. DM channel (middle), and DM channel vs. S/N (right). The panel on the bottom shows the location of events with S/N~$\geq 5$ in the DM-time plane. The two dispersed pulses occur at approximately $t = 42$ and $t = 49$ s, and have a distinct signature from radio interference, which spans an unrealistically large range of trial DM values (e.g. at $t = 4$ and $t = 62$ s).}
\label{rrat}
\end{figure}

Among the 28 pulsars discovered by PALFA so far, two objects fit the characteristics of a newly defined class of radio-loud neutron stars, Rotating RAdio Transients (RRATs, McLaughlin et al. 2006). J0628+09 was found only by the single pulse search algorithm; it exhibits sporadic bursts with a peak S/N of up to 100. Only 3 pulses were found in the 67 seconds of data of the discovery observation (Fig.\ref{rrat}), and the period was estimated to be 2.48~s. Longer follow-up observations allowed the pulsar to also be detected with the periodicity search algorithm and narrow down its period to 1.24~s. J1928+15 was also discovered only by its single pulses. In that case, only 2 bursts were detected during the 134 second observation, which was not enough to constrain the period. Follow-up observations of this object are still forthcoming. 

\subsection{Data Archive}

All raw data and data processing products from the PALFA survey will be archived permanently at a state-of-the-art tape library at Cornell Theory Center. This archive is expected to grow to $\sim 1$ PB by the time the survey is completed, and presents a challenge both with respect to database management and insuring data integrity and developing appropriate tools to make such a large archive publicly available online. Currently effort is under way to develop web services, web applications and a comprehensive online portal to the archive which will provide PALFA consortium members and external users with a means to request, download, and remotely manipulate raw data and data processing products. Through standardized web services and protocols, the PALFA survey portal will also be interfaced with and made available through the National Virtual Observatory.

\section{Future SKA Pulsar Survey}

With its unprecedented collecting area and sensitivity, the Square Kilometer Array (SKA) promises to revolutionize several areas of radio astronomy, including pulsar astrophysics. The frequency range of the SKA will be 0.1-25~GHz, with a bandwidth equal to $25\%$ of the observing frequency, to a maximum of 4 GHz bandwidth for frequencies above 16 GHz. It will be possible to observe simultaneously at two independent frequency bands with the same field-of-view center and 2 polarizations per band. $75\%$ of the collecting area will be within a 150~km diameter and the maximum baseline length will be up to 3000~km, allowing for an angular resolution of 0.1~arcseconds at 1.4~GHz (for full specification, see \tt http://www.skatelescope.org\rm).

The sensitivity of the SKA makes it especially attractive for pulsar searching. For a pulsar with a period $P$ and pulse width $W$ the minimum detectable flux density is 
\begin{equation}
S_{min} = \frac{m\sigma T_{sys}}{G\sqrt{n_p t_{int} \Delta\nu}}\left(\frac{W}{P-W}\right), 
\end{equation}
where $m$ is the number of $\sigma$ corresponding to the detection threshold, $T_{sys}$ is the system temperature, $G$ is the gain, $n_p$ is the number of polarizations summed, $t_{int}$ is the integration time and $\Delta\nu$ is the bandwidth. For the SKA, the factor $T_{sys}/G$ is $\sim 10$ times smaller than for the Arecibo telescope and $\sim 100$ times smaller than for the Green Bank telescope. Assuming $m = 8$, $T_{sys} = 25$~K, $t_{int} = 60$~s and $\Delta\nu = 0.5\nu$ gives $S_{min} \approx 1.4\mu$Jy (Kramer 2003). This corresponds to a minimum detectable luminosity of 0.1~mJy~kpc$^2$ for the Galactic Center region and 0.8~mJy~kpc$^2$ for the farther side of our Galaxy. A simulated SKA pulsar survey assuming an integration time of 10 minutes and all-sky coverage yields $\sim$~10000 new pulsar discoveries (Cordes, priv.comm.), amounting to a complete Galactic pulsar census. Among the most eagerly anticipated discoveries are pulsars closely orbiting the massive black hole in the Galactic Center. The high stellar density in this region means that similar to globular clusters, it will be a prime target for searching for binary and milliseconds pulsars, and through timing of these objects as they move through the potential of the black hole, we can hope to measure the black hole spin. The large number of projected SKA pulsar discoveries means that we will have probes of electron number density along many lines of sight through the Galaxy and will be able to refine models of ionized gas structure like the NE2001 model of Cordes \& Lazio (2002). In a similar manner, with the SKA we will be able to detect the brightest pulsars in nearby galaxies and use them as probes of the ionized intergalactic medium. Even though pulsars in other galaxies may not be detectable by periodic searches, young Crab-like pulsars can be detected by their giant pulses out to intergalactic distances. Currently our sample of known pulsars outside our own Galaxy is limited to $\sim 10$ objects in the Magellanic Clouds. With projected extragalactic pulsar discoveries by the SKA we can study the pulsar population statistics in different galaxy types. 

\begin{figure}
\centerline{\psfig{file=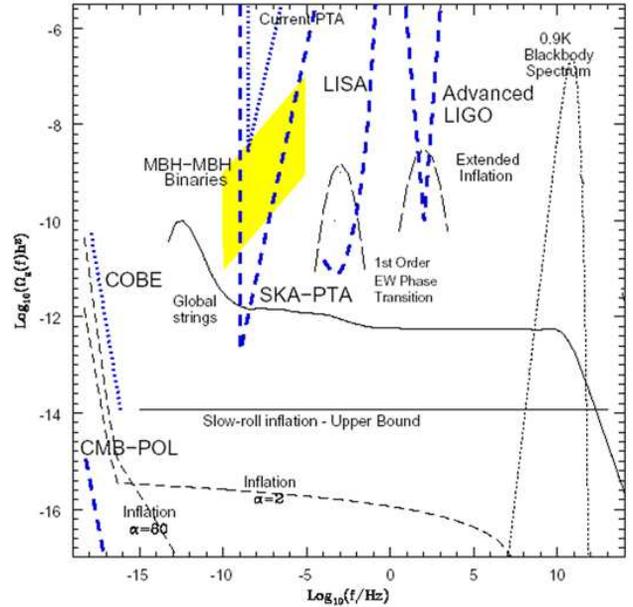,width=9.3cm,clip=}}
\caption{Comparison of expected sources of gravitational waves and sensitivity limits for various methods of their detection.}
\label{skapta}
\end{figure}

Another use of pulsars as probes is for gravitational wave detection. Multiple milliseconds pulsars found by the SKA can serve as the endpoints of arms of a huge gravitational wave detector: the Pulsar Timing Array (PTA). The concept for the PTA hinges on the fact that pulsars are extremely accurate natural clocks. A gravitational wave passing in the vicinity of the Earth will stretch and compress the space through which the pulsar signals have to travel in order to reach us. Correlated changes in the arrival times of pulses from pulsars in different areas of the sky will indicate space-time distortion due to gravitational waves. The dimensionless strain quantity $h_c\left(f\right)$ used to measure the magnitude of the distortion can be related to the observed scatter in arrival times and the integration time: $h_c\left(f\right) \sim \sigma_{TOA}/t_{int}$. Gravitational waves with frequencies on the order of nHz (nanoHertz) would be accessible only with the PTA (Fig.\ref{skapta}).

Last but not least, the large number of expected SKA pulsar discoveries means that it is statistically likely the sample will contain a corresponding number of exotic objects: relativistic binaries, double pulsar binaries, perhaps also the long sought-after sub-millisecond pulsars and pulsar-black hole binaries. The sample of binary pulsars whose masses can be determined can constrain the neutron star mass distribution. We do not yet know what are the physical limits on pulsar masses or periods. The large SKA pulsar sample would allow us to refine estimates of the likelihood of the slowest, fastest, least or most massive pulsar likely to exist, and to constrain the equation of state of matter under extreme conditions.

\begin{acknowledgements}
The PALFA survey is supported by the National Science Foundation through a cooperative agreement with Cornell University to operate the Arecibo Observatory. The survey has benefited from the expertise of staff at Arecibo, the National Astronomy and Ionospheric Center, and the Australia Telescope National Facility. We are grateful to Cornell Theory Center staff for their work in designing the PALFA database and porting of data processing code. The PALFA raw data archive and database work is supported by a National Science Foundation RI grant, a Microsoft E-Science grant, and by the Unisys Corporation.

The SKA development project is supported by scientific organizations and observatories in the participating countries (\tt http://www.skatelescope.org\rm).

\end{acknowledgements}
   


            \clearpage


\begin{thebibliography}{} 

\bibitem{}
Cordes, J. M. \& Lazio, T. J. W. 2002, astro-ph/0207156
\bibitem{}
Cordes J. M. et al., 2005, ApJ, 637, 446
\bibitem{}
Cordes, J. M. \& McLaughlin, M. A. 2003, ApJ, 596, 1142
\bibitem{}
Dowd, A., Sisk, W., \& Hagen, J. 2000, Astronomical Society of the Pacific Conference Series, 202, 275
\bibitem{}
Heiles, C. 2004, ALFA Memo, \tt http://alfa.naic.edu/memos \rm
\bibitem{}
Kramer, M. et al. 2004, New Astronomy Reviews, 48, 993
\bibitem{}
Lorimer, D. R. 2001, Arecibo Technical Memo No. 2001–01
\bibitem{}
Lorimer D. R. et al., 2006, ApJ, 640, 428
\bibitem{}
McLaughlin, M. A. et al. 2006, Nature, 439, 817

\end{thebibliography}
\end{document}